\documentclass[twocolumn,aps,epsfig,graphicx,showpacs]{revtex4}
\usepackage[dvips]{graphicx}
\newcommand{\be}{\begin{equation}}
\newcommand{\ee}{\end{equation}}

\newcommand{\bea}{\begin{eqnarray}}
\newcommand{\eea}{\end{eqnarray}}
\newcommand{\bd}{\begin{displaymath}}
\newcommand{\ed}{\end{displaymath}}
\newcommand{\bi}{\begin{itemize}}
\newcommand{\ei}{\end{itemize}}
\newcommand{\bc}{\begin{center}}
\newcommand{\ec}{\end{center}}
\newcommand{\bfl}{\begin{flushleft}}
\newcommand{\efl}{\end{flushleft}}
\newcommand{\bfr}{\begin{flushright}}
\newcommand{\efr}{\end{flushright}}
\newcommand{\f}{\frac}

\def\6{\partial}

\def\+{\!\!\!&&\!\!\!+~}
\def\-{\!\!\!&&\!\!\!-~}

\begin{document}
\title{ NMR  parameters in gapped graphene systems}
\author{Mircea Crisan}
\affiliation{Department of Theoretical Physics, ``Babe\c{s}-Bolyai" University, 400084 Cluj-Napoca, Romania}
\author{Ioan Grosu}
\affiliation{Department of Theoretical Physics, ``Babe\c{s}-Bolyai" University, 400084 Cluj-Napoca, Romania}
\author{Ionel \c{T}ifrea}
\affiliation{Department of Physics, California State University, Fullerton, CA 92831, USA}

\begin{abstract}
We calculate the nuclear spin-lattice relaxation time and the Knight shift for the case of gapped graphene systems. Our calculations consider both the massive and massless gap scenarios. Both the spin-lattice relaxation time and the Knight shift depend on temperature, chemical potential, and the value of the electronic energy gap. In particular, at the Dirac point, the electronic energy gap has stronger effects  on the system nuclear magnetic resonance parameters in the case of the massless gap scenario. Differently, at large values of the chemical potential, both gap scenarios behave in a similar way and the gapped graphene system approaches a Fermi gas from the nuclear magnetic resonance parameters point of view. Our results are important for nuclear magnetic resonance measurements that target the $^{13}$C active nuclei in graphene samples.
\end{abstract}
\maketitle

\section{Introduction}

Graphene, a two-dimensional (2D) carbon based material, has attracted a lot of research interest due to its unconventional physical properties and its various practical applications  \cite{novoselov1,novoselov2}.   

The carbon atoms in graphene are arranged in a hexagonal, honeycomb structure that can be seen as a triangular lattice with two atoms per unit cell. Of major importance are the two inequivalent points ($K$ and $K'$) at the corners of the graphene Brillouin zone. The electronic properties of monolayer graphene are those of a semimetal and they are due to a band structure that exhibits a Fermi surface that is reduced to the two points $K$ and $K'$ (the so-called Dirac points) with a zero density of states at the Fermi energy. The low-energy excitations around  $K$ ($K'$) are characterized by quasiparticles that follow a linear dispersion in the vicinity of the Fermi points \cite{wallace}
\begin{equation}\label{dispersion}
E_\pm(k)=\pm  v_F k + O((k/K)^2)\;,
\end{equation}
with Dirac fermions moving at a speed $v_F$ about 300 times smaller than the speed of light \cite{castroneto,dassarma}. Consequently, many of the physical properties of graphene are similar to properties characteristic to the physics of quantum electrodynamics, although at much lower speed. In particular, when subject to external magnetic fields graphene exhibits an anomalous integer quantum Hall effect that can be observed at room temperature \cite{gusynin,peres,novoselov3}, a phenomena that is strictly related to the Dirac fermions behavior.

From the applications point of view, the 2D graphene is a very promising high quality material. As mentioned, graphene has a linear, gapless, energy spectrum and exhibits a metallic-like electric conductivity even in the limit of low carrier concentration. However, despite graphene's outstanding physical properties, most electronic applications require the presence of a gap between the conduction and the valence bands. One possibility to open a gap in the graphene's energy spectrum is to use geometry confinement as in quantum dots or nanoribbons \cite{geim,trauzettel,nakada,brey,chen,han}. Differently, Zhou {\em et al.}  \cite{zhou1,zhou2} proved that when graphene is epitaxially grown on SiC substrate a gap of about $2\Delta \sim0.26$ eV opens in its energy spectrum, leading to a dispersion relation (massive gapped spectrum)
\begin{equation}\label{massive}
E^{ms}_{\pm}(k)=\pm\sqrt{(v_F k)^2+\Delta^2}\;.
\end{equation}
Angle resolved photoemission spectroscopy (ARPES)  in epitaxially grown graphene proved that the magnitude of the band-splitting gap decreases as the number of layers in the system increases \cite{zhou1}. Similar ARPES results were reported by Bostwick {et al.} \cite{bostwick1,bostwick2}, although the explanation for the nature of the energy gap differs. In particular, the dispersion relation for the massive gaped spectrum (see Eq. (\ref{massive})) cannot describe correctly the electronic dispersion far from the Dirac point. Benfatto and Cappelluti \cite{benfatto} proposed a phenomenological gap scenario that aims to reconcile the gapped nature of the system's electronic spectrum and the massless character of the fermions in graphene. In this second scenario (massless gapped spectrum), the electron dispersion relation is given by 
\begin{equation}\label{massless}
E^{ml}_{\pm}(k)=\pm(v_F k+\Delta)\;.
\end{equation}
The massless gapped scenario seems to correctly account for spectroscopic measurements in epitaxially grown graphene.

Low dimensional carbon systems that are based on graphene structures such as graphene monolayers and bilayers, carbon nanotubes and ribbons, or graphene quantum dots, all display a variety of interesting properties that are intrinsically due to Dirac fermions \cite{castroneto,dassarma}. In general, the linear or gapped graphene electronic spectrum leads to significant deviations from the standard Fermi liquid properties of a metal. For example, using $^{13}$C nuclear magnetic resonance (NMR), Singer {\em et al.} found evidence for a gap opening in the electronic excitation spectrum of carbon nanotubes \cite{singer}. Additionally, the ratio $1/T_1 T$ ($T_1$ - is the nuclear spin relaxation time and $T$ is the temperature) follows the expected Korringa law \cite{slichter} at high temperatures, but consistently deviates from the standard metallic behavior in the low temperature limit \cite{singer}. Dora {\em et al.} explained the anomalous behavior of the NMR data in carbon nanotubes using a combination of the Luttinger liquid and Luther-Emery liquid theories (interacting one dimensional electronic systems without and with a gap, respectively) \cite{dora1}.

The NMR investigation of carbon based systems is difficult as only the isotope $^{13}$C (natural abundance 1.1\%) has a non-zero nuclear spin, therefore the need for special graphene samples enriched with this particular isotope \cite{cai}. Although the hyperfine interaction was predicted to be of the order of 1$\mu$eV and highly anisotropic \cite{yazyev, fischer}, experimental measurements in carbon nanotubes found its value about two orders of magnitude higher around 100 $\mu$eV \cite{churchill}. Also, the local hyperfine magnetic field in magnetic carbon based materials was estimated to be of the order of 18 - 21 T \cite{freitas}.

Theoretically, NMR in graphene samples was investigated by several authors \cite{dora2,ma,frota}. Dora and Simon calculated the Knight shift and the spin-lattice relaxation time for a single sheet of carbon atoms and identified three possible regimes for the system: Fermi gas, Dirac gas, and an extreme quantum limit behavior \cite{dora2}.  Similar deviations from the standard metallic behavior were reported for NMR in graphene samples using a tight-binding model for both single layer systems \cite{ma} and bilayer systems \cite{frota}.

Here, we provide an analytical investigation of the NMR Knight shift and spin-lattice relaxation time for the case of a gapped graphene system. Our analysis includes both the massive and massless gapped spectrum scenarios. We specifically discuss the influence of the gap on the system's NMR properties and analyze the possible crossover between Dirac and Fermi liquid models. The paper is organized as follows: In section II we discuss the system's density of states for massive and massless fermions. In Section III we present results for the system's NMR properties and we discuss the possible crossover between the Dirac and Fermi liquid models. Finally, in the last section we present our conclusions.

\section{Theoretical model}

Single layered graphene systems  in the low energy limit can be described by the two dimensional Dirac  hamiltonian
\begin{equation}\label{singlelayerhamiltonian}
H_0=v_F(\sigma_x p_x+\sigma_y p_y)\;,
\end{equation}
where $v_F$ is the Fermi (Dirac) velocity, $\bf{k}$ is the momentum relative to the K (K') point, and $\sigma_i$ are the 2$\times$2 Pauli matrices. The hamiltonian's  corresponding eigenvalues are given by Eq. (\ref{dispersion}) and the system's density of states has a simple linear form \cite{castroneto}:
\begin{equation}\label{DOSgraphene}
\rho(E)=\rho_0|E|\;,
\end{equation}
where $\rho_0=A_c/2\pi v_F^2$ ($A_c$ is the unit cell area). 

\begin{figure}[t]
\centering \scalebox{0.95}[0.95]{\includegraphics*{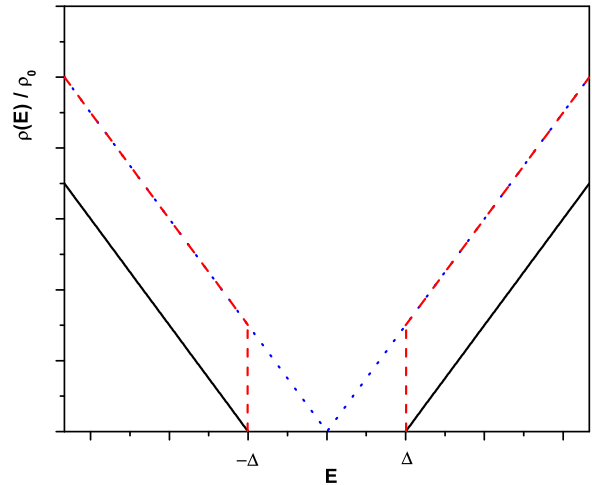}}
\caption{(Color online) Electronic density of states as function of energy for pure graphene (blue line), gapped graphene in the massive gapped scenario (red line), and gapped graphene in the massless gapped scenario (black line).}
\label{fig1}
\end{figure}

To account for a gap opening in the electronic spectrum of graphene, an additional term 
\begin{equation}\label{gaphamiltonian}
H_1\propto \Delta\sigma_z\;
\end{equation}
has to be included in the system's hamiltonian. Usually, this term is related to an inequivalence between the A and B sublattices of graphene \cite{zhou1}. Theoretically, Manes {\em et al.} proved that a similar gap can open when a translationally invariant perturbation acts on the graphene system \cite{manes}. The main effect due to this additional term  in the hamiltonian is the change in the system's electronic energy dispersion relation, in this case the Dirac fermions spectrum being well described by Eq. (\ref{massive}). Accordingly, the system's density of states has a slightly different form \cite{koshino,ludwig} 
\begin{equation}\label{DOSmassive}
\rho_{ms}(E)=\rho_0|E|\Theta(E^2-\Delta^2)\;,
\end{equation}
with $\Theta(x)$ being the standard step function. In this case, in the low energy limit, the graphene electrons acquire a finite mass
\begin{equation}\label{massiveapprox}
E^{ms}_{\pm}(k)\simeq \pm\left(\Delta+\f{k^2}{2 m_{eff}}\right)\;,
\end{equation}
with the effective electron mass $m_{eff}=\Delta^2/v_F^2$. The massive gapped scenario was used to describe ARPES data \cite{zhou1}, although it is still controversial if the model correctly describe the experimental data. In particular, far from the Dirac points, the model fails to correctly describe the shape of the ARPES curves. Additional differences between the theoretical model and the experimental data appear also in the shape of the dispersion spectrum at finite $k$ (away from $K$ ($K'$) point), when instead of the predicted parabolic behavior, the experimental ARPES data show more of a linear, massless, dispersion.

In the second scenario, Benfatto and Cappelluti \cite{benfatto} introduced a gap in graphene's electron spectrum using a phenomenological structure for the system's self-energy, but at the same time conserving the massless characteristics of the Dirac electrons (see Eq. (\ref{massless})). In this case the corresponding density of states can be evaluated to be
\begin{equation}\label{DOSmassless}
\rho(E)=\rho_0\left(|E|-\Delta\right)\Theta(|E|-\Delta)\;.
\end{equation}

In Figure \ref{fig1} we present the electronic density of states for single-layered graphene (blue line) and gapped graphene in the two scenarios: massive gapped scenario (red line) and massless gapped scenario (black line). Note that in the massless gapped scenario the system's density of states does not extrapolate to zero at the Dirac point.

\section{NMR parameters}

NMR is one of the most developed experimental techniques with wide applications in many scientific fields. In NMR experiments, an external magnetic field is used to polarize nuclei across the investigated sample, and thereafter, using a perpendicular variable magnetic field we can induce transitions in the nuclear system and measure resonance frequencies \cite{slichter}. There are two important parameters that can be measured in NMR spectroscopy, i.e., the nuclear spin-lattice relaxation time $T_1$ and the shifts of the NMR resonances. Various types of interactions can influence these parameters \cite{slichter}.  In a metal, the hyperfine interaction between nuclear and electronic spins described by the standard Fermi contact term, leads to a temperature independent shift in the NMR spectrum resonances (Knight shift) and to the well known Korringa relation, $1/(T_1T)\sim constant$ \cite{slichter}.Dora and Simon investigated the unusual hyperfine interaction in graphene monolayer systems and found strong deviations for the graphene NMR parameters compare to the corresponding ones in standard metals \cite{dora2}. In graphene monolayer systems the Fermi contact term in the electron-nuclear spin interaction Hamiltonian differs significantly from the corresponding term in standard metals: ${\bf r}\times {\bf j}$ replaces the usual term ${\bf r}\times {\bf p}$. Note that in graphene systems ${\bf j} \sim {\bf \sigma}$ (${\bf \sigma}$ is a vector of the Pauli matrices), and therefore all the changes in the NMR parameters behavior. Additionally, Dora and Simon found that the systems is characterized by  three different regimes: Fermi gas, Dirac gas, and extreme quantum limit, with NMR parameters behaving differently in each of these limits \cite{dora2}. Frota and Ghosh reported similar deviations from the standard Korringa relation when they studied the main NMR parameters for graphene samples in the framework of a tight-binding model \cite{frota}. The role of the hyperfine interaction was also considered in carbon based nanotubes and quantum dots \cite{palyi,csiszar}.

The spin-lattice relaxation time and the Knight shift can be calculated as \cite{slichter}
\begin{equation}\label{T1general}
\frac{1}{T_{1}T} = \frac{C^{2} \pi k_{B}}{\hbar}\int_{-\infty}^{\infty} \frac{[\rho (\omega)]^{2}d\omega}{4k_BT\cosh^{2}\left[(\omega-\mu)/2k_BT\right]}
\end{equation}
and
\begin{equation}\label{Knightshift}
K= \frac{A\gamma_{e}}{2\gamma_{n}}\int_{-\infty}^{\infty} \frac{\rho (\omega)d\omega}{4k_BT\cosh^{2}\left[(\omega-\mu)/2k_BT\right]}\;.
\end{equation}
Above, $k_B$ is the Boltzmann constant, $T$ is the temperature, $\gamma_e$ is the electron gyromagnetic ratio, $\gamma_n$ is the nuclear gyromagnetic ratio, $\mu$ is the electron chemical potential, and $A$ is  the hyperfine constant. The other constant,  $C$, is defined using both the hyperfine constant $A$ and the orbital interaction constant, $J_{orb}$ \cite{dora2}. In particular, both parameters, the spin-lattice relaxation time and the Knight shift, should be anisotropic, however, an argument can be made that due to the orbital interaction in graphene both parameters can be considered isotropic \cite{dora2,yazyev}. 

\begin{figure}[b]
\centering \scalebox{0.95}[0.95]{\includegraphics*{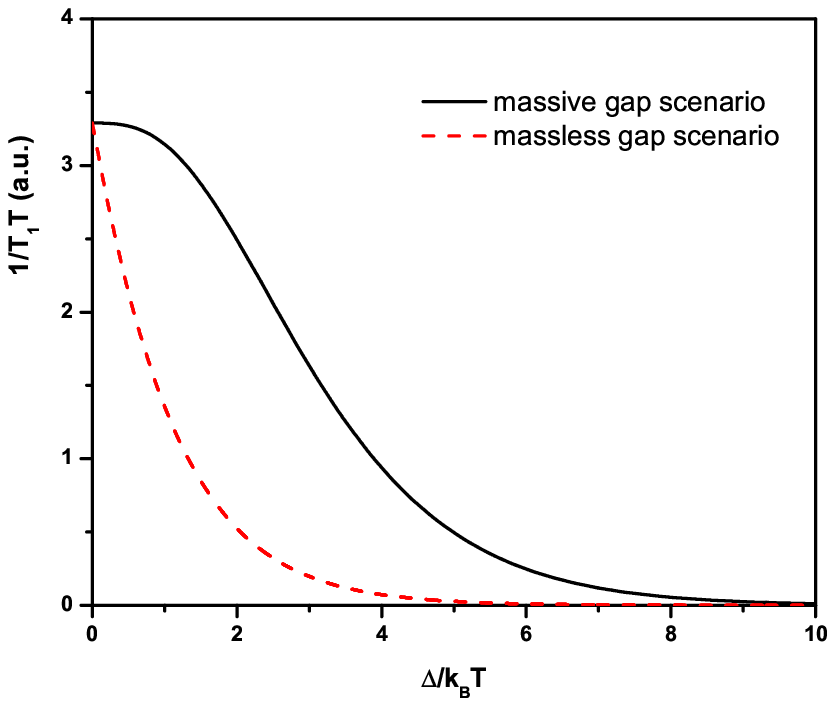}}
\caption{(Color online) The $1/T_1T$ ratio as function of $\Delta/k_BT$ for gapped graphene close to the Dirac point ($\mu/k_BT\rightarrow 0$) (black line - massive gap scenario; red line - massless gap scenario).}
\label{fig2}
\end{figure}

In the following, we will estimate the NMR parameters for both the massive gapped and massless gap scenarios in a two dimensional graphene system. Equations (\ref{T1general}) and (\ref{Knightshift}) are expected to hold even in the case of a gapped electronic system. The standard procedure involved in the derivation of the electron-nuclear spin interaction Hamiltonian requires the replacement of the electron momentum ${\bf p}$ with the generalized momentum ${\bf p}+e{\bf A}$, where ${\bf A}$ is the vector potential due to the magnetic field produced by the nuclear spin. As the energy gap in the electron spectrum is momentum independent, its presence will not influence the form of the hyperfine interaction, and accordingly the standard equations for the nuclear spin relaxation time and Knight shift will hold.

\subsection{The spin-lattice relaxation time}

The spin-lattice relaxation time in the case of single-layered graphene samples behaves as $1/(T_1T)\sim \textrm{max}\;[\mu^2,T^2]$, so its temperature dependence is relevant only close to the Dirac point (small chemical potential $\mu$). The situation is different in the presence of impurities, when the system's density of states changes and the spin-lattice relaxation time reproduces the Fermi-gas behavior \cite{sharapov,dora2}.

We can calculate the spin-lattice relaxation time for the gapped graphene system in the two different scenarios using Eq. (\ref{T1general}) along with Eqs. (\ref{DOSmassive}) and (\ref{DOSmassless}). In the case of the massive gap scenario one finds:
\begin{widetext}
\begin{eqnarray}\label{T1massive}
\f{1}{T_1T}=\f{C^2\pi k_B\rho_0^2}{\hbar}&&\left\{\mu^2+\f{(\pi k_BT)^2}{3}+\Delta^2\left(\f{1}{1+\exp{\left[-\f{\mu_-}{k_BT}\right]}}-\f{1}{1+\exp{\left[-\f{\mu_+}{k_BT}\right]}}\right)\right.\nonumber\\
&&\left.+2k_BT\Delta\left[\ln{\left(1+\exp{\left[\f{\mu_+}{k_BT}\right]}\right)}+\ln{\left(1+\exp{\left[\f{\mu_-}{k_BT}\right]}\right)}\right]\right.\nonumber\\
&&\left.+2\left(k_BT\right)^2\left[Li_2\left(1+\exp{\left[\f{\mu_+}{k_BT}\right]}\right)-Li_2\left(1+\exp{\left[\f{\mu_-}{k_BT}\right]}\right)\right]\right\}\;,
\end{eqnarray}
where $\mu_\pm=\mu\pm\Delta$, and 
\begin{equation}\label{dilog}
Li_2(x)=-\int_0^x dt\f{\ln{(1-t)}}{t} 
\end{equation}
is the dilogarithm function. On the other hand, in the case of the massless gap scenario we have
\begin{eqnarray}\label{T1massless}
&&\f{1}{T_1T}=\f{C^2\pi k_B\rho_0^2}{\hbar}\left\{\mu_+^2+\f{(\pi k_BT)^2}{3}+2\left(k_BT\right)^2\left[Li_2\left(1+\exp{\left[\f{\mu_+}{k_BT}\right]}\right)-Li_2\left(1+\exp{\left[\f{\mu_-}{k_BT}\right]}\right)\right]\right\}.
\end{eqnarray}
\end{widetext}
Note that for both case scenarios in the limit $\Delta\rightarrow 0$ we recover the previous result for single-layered two dimensional graphene systems \cite{dora2}
\begin{equation}\label{T1graphene}
\f{1}{T_1T}=\f{C^2\pi k_B\rho_0^2}{\hbar}\left[\mu^2+\f{(\pi k_BT)^2}{3}\right]\;.
\end{equation}

Figure \ref{fig2} presents the spin-lattice relaxation time as function of the gap magnitude around the Dirac point ($\mu\rightarrow 0$ limit). The spin-lattice relaxation time behavior is very similar in the two scenarios. In the small gap limit, $\Delta/k_BT\ll 1$, the spin-lattice relaxation time at the Dirac point behaves similar to the pure graphene case, i.e., $T_1\sim T^{-3}$, and the nuclear spins are not relaxed by conduction electrons at $T=0$. In the large gap limit, $\Delta/k_BT\gg 1$, $1/T_1\rightarrow 0$ even at finite temperatures, and the presence of an energy gap in graphene's electronic spectrum means that the nuclear spins will not be relaxed by conduction electrons as long as the gap is large enough.

\subsection{The Knight shift}

The Knight shift for single-layered  two dimensional graphene samples behaves as $K\sim \textrm{max}\;[|\mu|,T]$, or as $K\sim \Gamma \ln{(D/\Gamma)}$ in the presence of impurities \cite{dora2} ($\Gamma$ is the scattering rate and $D$ is the cutoff in the continuum theory).

In the case of the gapped graphene, the Knight shift can be evaluated using Eq. (\ref{Knightshift}) and Eqs.  (\ref{DOSmassive}) and (\ref{DOSmassless}). In the massive gap scenario we find:
\begin{widetext}
\begin{eqnarray}\label{Kmassive}
K=\f{A\gamma_e\rho_0}{2\gamma_n}&&\left\{-\mu+\Delta\left[\f{1}{1+\exp{\left[-\f{\mu_-}{k_BT}\right]}}-\f{1}{1+\exp{\left[-\f{\mu_+}{k_BT}\right]}}\right]\right.\nonumber\\
&&\left.+k_BT\left[\ln{\left(1+\exp{\left[\f{\mu_+}{k_BT}\right]}\right)}+\ln{\left(1+\exp{\left[\f{\mu_-}{k_BT}\right]}\right)}\right]\right\}\;.\nonumber\\
\end{eqnarray}
In the massless gap scenario the Knight shift is given by
\begin{eqnarray}\label{Kmassless}
K=\f{A\gamma_e\rho_0}{2\gamma_n}\left\{-\mu_++k_BT\left[\ln{\left(1+\exp{\left[\f{\mu_+}{k_BT}\right]}\right)}+\ln{\left(1+\exp{\left[\f{\mu_-}{k_BT}\right]}\right)}\right]\right\}\;.
\end{eqnarray}
\end{widetext}
Again, in the limit $\Delta\rightarrow 0$ the previous result for single-layered graphene systems is recovered \cite{dora2}
\begin{equation}\label{Kgraphene}
K=\f{A\gamma_e\rho_0k_BT}{\gamma_n} \ln{\left[2\cosh{\left(\f{\mu}{2k_BT}\right)}\right]}\;.
\end{equation}

\begin{figure}[b]
\centering \scalebox{0.95}[0.95]{\includegraphics*{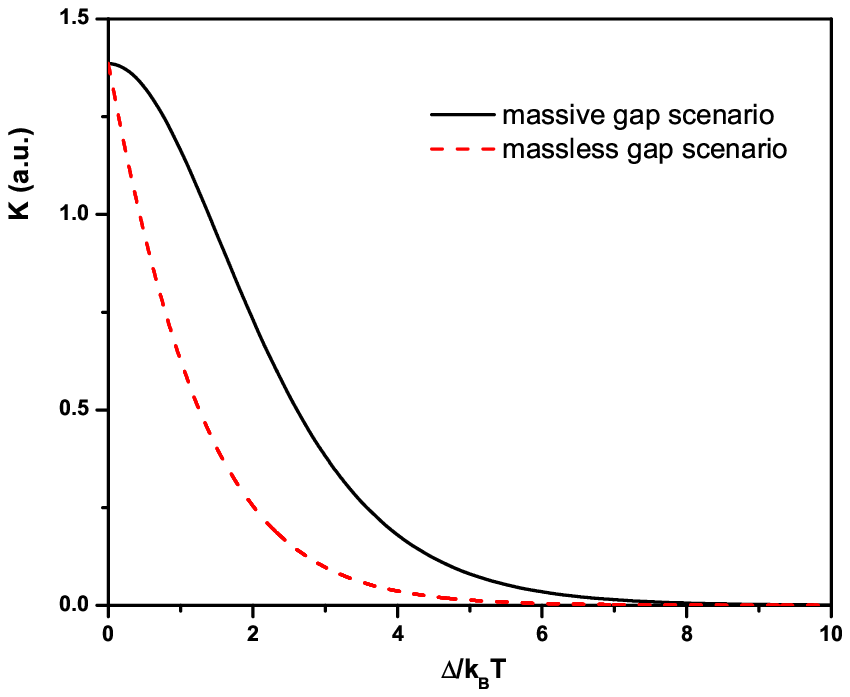}}
\caption{(Color online) The Knight shift $K$ as function of $\Delta/k_BT$ for gapped graphene close to the Dirac point ($\mu/k_BT\rightarrow 0$) (black line - massive gap scenario; red line - massless gap scenario).}
\label{fig3}
\end{figure}

Figure \ref{fig3} presents the Knight shift as function of the gap magnitude around the Dirac point. As a general result the Knight shift cancels in the  large gap limit for both case scenarios.

Standard Fermi gases obey the Korringa relation, i.e., the ratio $R_F=(1/T_1TK^2)_F$ is a constant ($R_F=4\pi k_B(\gamma_n/\gamma_e)^2/\hbar$). Dora and Simon \cite{dora2} analyzed this ratio for pure two-dimensional graphene and found that based on the value of the chemical potential the system is subject to a crossover from a Dirac gas at low chemical potential values, $\mu/k_BT\ll1$, to a Fermi gas at high chemical potential values, $\mu/k_BT\gg 1$. In particular, for a Dirac gas, the Korringa ratio
\begin{equation}\label{RDirac}
\f{1}{T_1TK^2}=\left(\f{1}{T_1TK^2}\right)_F\left(\f{C^2}{A^2}\right) F\left(\f{\mu}{k_BT}\right)\;,
\end{equation}
with 
\begin{equation}\label{ScaleFunction}
F(x)=\f{3x^2+\pi^2}{12\ln^2{\left[2\cosh{\left(x/2\right)}\right]}}.
\end{equation}
In the Dirac gas limit, at the Dirac point, $F(0)$=1.71, although in the opposite limit, $\mu/k_BT\rightarrow\infty$, $F(\infty)=1$ and the Fermi gas behavior is recovered. 

\begin{figure}[b]
\centering \scalebox{0.95}[0.95]{\includegraphics*{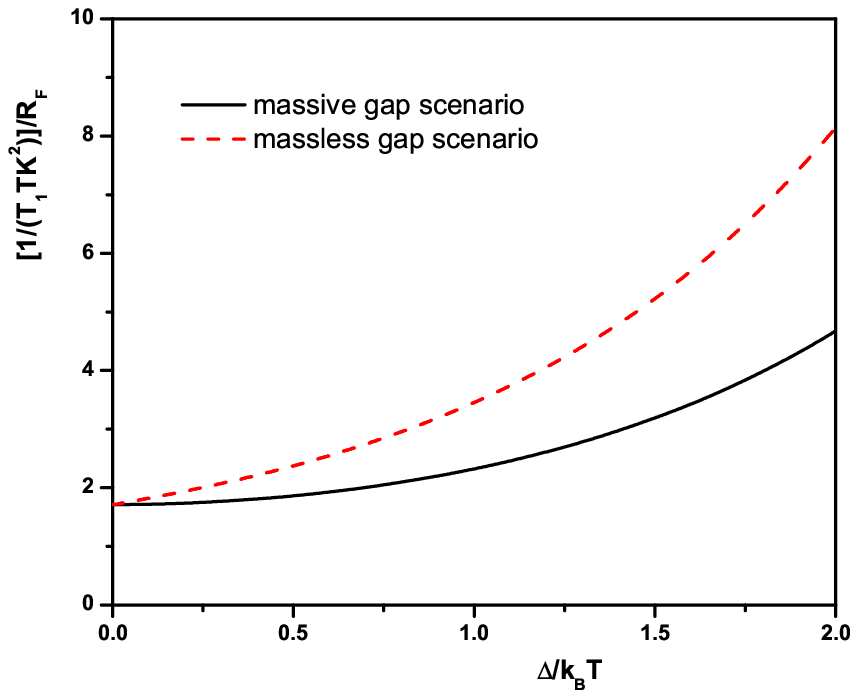}}
\caption{(Color online) The Korringa ratio as function of $\Delta/k_BT$ for gapped graphene close to the Dirac point ($\mu/k_BT\rightarrow 0$) (black line - massive gap scenario; red line - massless gap scenario).}
\label{fig4}
\end{figure}

In a gapped graphene system, the same ratio depends both on the chemical potential and on the size of the energy gap
\begin{equation}
\f{1}{T_1TK^2}=\left(\f{1}{T_1TK^2}\right)_F\left(\f{C^2}{A^2}\right) G\left(\f{\mu}{k_BT},\f{\Delta}{k_BT}\right)\;,
\end{equation}
where $G(x,y)$ is a general function who's exact form depends on the gap scenario (massive vs massless) we considered. Strictly at the Dirac point, $\mu/k_BT=0$, the value $G(0,\Delta/k_BT)$ is higher for the massless gap scenario, with the exception of $G(0,0)$, where the two gap scenarios give the same value for the Korringa ratio, $G(0,0)$=1.71. Figure \ref{fig4} highlights the dependence of the Korringa ratio on the value of the energy at the Dirac point for the two possible scenarios.

Figure \ref{fig5} presents the Korringa ratio as function of the chemical potential $\mu/k_BT$ for different values of the electronic energy gap, $\Delta/k_BT$, for the massive gap scenario (Fig. \ref{fig5}a) and the massless gap scenario (Fig. \ref{fig5}b), respectively. As we already discussed, the Korringa ratio is bigger in the case of the massless gap scenario for the same value of the energy gap when the system approaches the Dirac point, $\mu/k_BT\ll 1$. On the other hand, on the opposite limit $\mu/k_BT\gg 1$, both scenarios are characterized by the same Korringa ratio, $1/T_1TK^2=R_F$. In other words, at large chemical potential values, the gapped graphene system behaves as a regular Fermi gas, regardless the gap scenario we consider.

\section{Conclusions}

\begin{figure}[t]
\centering \scalebox{0.75}[0.75]{\includegraphics*{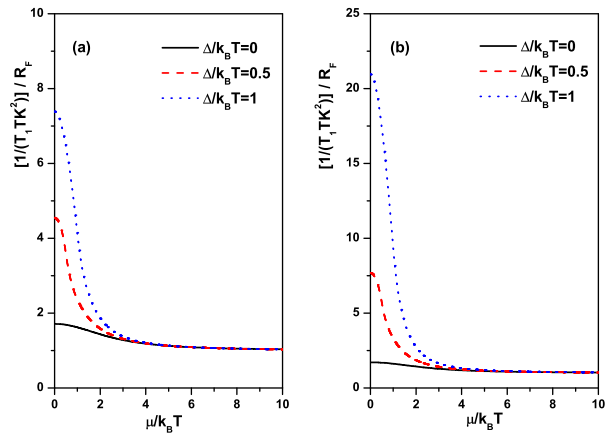}}
\caption{(Color online) The Korringa ratio  as function of the chemical potential $\mu/k_BT$  for different values of the energy gap  (black line - $\Delta/k_BT=0$; red line - $\Delta/k_BT=0.5$; blue line - $\Delta/k_BT=1$) in the two gap scenarios ((a) - massive gap scenario; (b) - massless gap scenario).}
\label{fig5}
\end{figure}

In summary we considered the NMR parameters for the case of gapped graphene systems. The possibility of a gap opening in the electronic spectrum of graphene samples was studied within two scenarios: the massive gap scenario \cite{zhou1} and the massless gap scenario \cite{benfatto}. Although both scenarios are phenomenological in nature, some theoretical arguments support their assumptions. In the first case, the inequivalence of graphene's two sublattices $A$ and $B$ can break the system's symmetry and leads to a modified electronic energy spectrum \cite{zhou1,manes}. On the other hand, in the second case, the gap is similar to an order parameter and is the consequence of many-body effects \cite{cappeluti}.

 Although within the reach of NMR spectroscopy, graphene samples suffer from the lack of active nuclei, as only the $^{13}$C isotope has a non-zero nuclear spin value. Therefore, NMR in graphene samples was demonstrated for $^{13}$C specially enriched samples \cite{cai}. On the other hand, from the theoretical point of view, several studies analyzed the nuclear spin-lattice relaxation time and the Knight shift related to the electron - nuclear spin interaction in graphene samples \cite{dora2,csiszar,palyi,frota}.

We considered the spin-lattice relaxation time and the Knight shift for the case of gapped graphene. Our analytical results, obtained for both massive and massless gap scenarios, are valid for the case when Landau levels can be neglected in graphene's electronic spectrum. Within both gap scenarios, the spin-lattice relaxation time and the Knight shift are strongly influenced by the presence of an energy gap in the electronic spectrum of the system. For both case scenarios we recover previous results for the spin-lattice relaxation time and the Knight shift obtained for single-layered graphene in the absence of the energy gap $\Delta\rightarrow 0$. For finite energy gaps, in particular at the Dirac point ($\mu\rightarrow 0$), the massless gap scenario is showing a stronger dependence on the value of the gap-temperature ratio for both the spin-lattice relaxation time and Knight shift.  At a fixed temperature, for small gap values, $\Delta/k_BT\ll 1$, $T_1\sim T^{-3}$, and the nuclear spins are not relaxed by conduction electrons when $T\rightarrow 0$. On the other hand, for large gap values, $\Delta/k_BT\gg 1$, $1/T_1\sim 0$, and the nuclear spins are not relaxed by conduction electrons regardless the value of the system's temperature. In the low gap limit, at fixed temperatures, the Knight shift appears to be larger in the case of the massive gapped scenario. In the opposite limit, for large values of the electron gap, for both case scenarios the Knight shift cancels.

The Korringa ratio, $1/T_1TK^2$, can be used to estimate the metallic character of the system. In the case of a Fermi gas this ratio is a constant. Our analytical calculations proved that in the case of the gapped graphene system the Korringa ratio is a function of the system's chemical potential, the value of the electronic gap, and temperature. At fixed temperatures, close to the Dirac point $\mu/k_BT\rightarrow 0$, the Korringa ratio is finite and increases faster as function of the electronic gap in the massless gap scenario. As function of the system's chemical potential, $\mu/k_BT$, the Korringa ratio approaches the Fermi gas value in the limit $\mu/k_BT\gg 1$, independent of the value of the energy gap. 

Based on the system's NMR parameters one can discuss the nature of electronic system in gapped graphene, i.e., Dirac gas vs. Fermi gas. Figure \ref{fig5} suggests that the system behaves as a Dirac gas for all values of the energy gap (including $\Delta/k_BT=0$) in the limit of a low chemical potential $\mu/k_BT\ll 1$. On the other hand, at large chemical potential values $\mu/k_BT\gg 1$ the system behaves as a Fermi gas for all values of the energy gap. Figure \ref{fig5} suggests that the crossover from the Dirac gas to the Fermi gas is almost independent of the energy gap value for both considered scenarios. In general, the chemical potential value can be controlled by doping or by external voltages. Differently, the crossover between the Dirac and Fermi gasses, can be also triggered by temperature if one fixes the values of the chemical potential and the energy gap in the system.

\end{document}